\documentclass[twocolumn,preprintnumbers, superscriptaddress, reprint, longbibliography, prl]{revtex4-1}
\usepackage{amsmath}
\usepackage{stmaryrd}
\usepackage{txfonts}
\usepackage{amssymb}
\usepackage{mathrsfs}
\usepackage{graphicx}
\usepackage{dcolumn}
\usepackage{bm}
\usepackage{braket}
\usepackage{epsfig}
\usepackage{color}
\usepackage{ulem}
\usepackage{bibunits}

\usepackage[colorlinks=true, linkcolor=blue, citecolor=blue, urlcolor=blue]{hyperref}
\begin{document}
\title{Kondo enabled transmutation between spinons and superconducting vortices: origin of magnetic memory in 4Hb-$\mathrm{TaS_2}$ }
\author{Shi-Zeng Lin}

\affiliation{Theoretical Division, T-4 and CNLS, Los Alamos National Laboratory, Los Alamos, New Mexico 87545, USA}
\affiliation{Center for Integrated Nanotechnologies (CINT), Los Alamos National Laboratory, Los Alamos, New Mexico 87545, USA}

\begin{abstract}
Recent experiments [Persky {\it{et al.}}, Nature {\bf{607}}, 692 (2022)] demonstrate a magnetic memory effect in 4Hb-$\mathrm{TaS_2}$ above its superconducting transition temperature, where Abriokosov vortices are generated spontaneously by lowering temperature at zero magnetic field after field training the normal state. Motivated by the experiment, we propose the chiral quantum spin liquid (QSL) stabilized in the constituent layers of 4Hb-$\mathrm{TaS_2}$  as a mechanism. We model 4Hb-$\mathrm{TaS_2}$ as coupled layers of the chiral QSL and superconductor. Through the Kondo coupling between the localized moments and conduction electrons, there is mutual transmutation between spinons and vortices during the thermal cycling process, which yields magnetic memory effect as observed in experiments. We also propose a mechanism to stabilize the chiral and nematic superconductivity in 4Hb-$\mathrm{TaS_2}$ through the Kondo coupling of conduction electrons to the chiral QSL. Our results suggest 4Hb-$\mathrm{TaS_2}$ as an exciting platform to explore the interplay between QSL and superconductivity through the Kondo effect.    

\end{abstract}
\date{\today}
\maketitle

Quantum spin liquid (QSL) is an exotic  state of matter, where electron spin fractionalizes into more elementary degree of freedom that interacts through a dynamical gauge field. ~\cite{Balents2010,Savary_2016,RevModPhys.89.025003,Takagi2019} The existence of QSL has been well established by the exactly solvable models. Nevertheless, unambiguous experimental identification of the QSL remains a huge challenge despite many encouraging signs have been detected. The QSL can serve as a mother state to induce other novel quantum states. For instance, one can obtain unconventional superconductivity by doping QSL~\cite{Anderson1973} or by coupling QSL to metals through the Kondo coupling. \cite{Coleman_1989}

The recent observation of a magnetic memory and spontaneous vortices in a van der Waals superconductor 4Hb-$\mathrm{TaS_2}$ suggests the possible existence of QSL in this compound. \cite{Persky2022} 4Hb-$\mathrm{TaS_2}$ consists of two alternatingly stacked layers of octahedral $\mathrm{TaS_2}$ (1T-$\mathrm{TaS_2}$) and trigonal prismatic $\mathrm{TaS_2}$ (1H-$\mathrm{TaS_2}$). Both 1T-$\mathrm{TaS_2}$ and 1H-$\mathrm{TaS_2}$ can exist in a bulk form. The 1T-$\mathrm{TaS_2}$ bulk was argued to host QSL. \cite{Law_Lee_2017} The 1T-$\mathrm{TaS_2}$ undergoes an incommensurate charge density wave (CDW) transition around 350 K, followed by another transition to a commensurate CDW around 200 K, forming a $\sqrt{13}\times \sqrt{13}$ structure. \cite{Wilson1975} The unit cell is enlarged to having 13 Ta ions, where each Ta ions contributes one 5d electron. The unit cell forms a triangular lattice. The observed insulating behavior implies a Mott insulating state in 1T-$\mathrm{TaS_2}$ below 200 K. Indeed, the lower and upper Hubbard bands have been observed by scanning tunneling microscope. \cite{PhysRevX.7.041054,Vano2021}. However, no magnetic order and even the formation of localized moment has been observed down to the lowest temperature that is much smaller than the estimated exchange coupling between localized spins. \cite{PhysRevB.96.195131,Klanjsek2017,Manas2021,PhysRevB.102.054401} These experiments support the existence of a QSL, either a fully gapped $Z_2$ QSL or a Dirac QSL, in 1T-$\mathrm{TaS_2}$ proposed by Law et al. \cite{Law_Lee_2017}. Later, a more refined modeling  calculations based a spin Hamiltonian that is appropriate for 1T-$\mathrm{TaS_2}$ concludes a QSL with spinon Fermi surface \cite{PhysRevLett.121.046401}. The QSL picture is also supported by other measurements. \cite{PhysRevResearch.2.013099,Vano2021,Ruan2021}

On the other hand, 2H-$\mathrm{TaS_2}$ (two layers of 1H-$\mathrm{TaS_2}$) is a superconductor with $T_c=0.7$ K. \cite{Nagata_Aochi1992} Therefore 4Hb-$\mathrm{TaS_2}$ offers an exciting platform for studying the interplay between superconductivity and QSL. One expects a Kondo coupling between the metallic layer 1H-$\mathrm{TaS_2}$ and the Mott insulator 1T-$\mathrm{TaS_2}$, which has been confirmed through the observation of the Kondo resonance peak by scanning tunneling microscopy. \cite{Vano2021,Ruan2021} \footnote{The experiment in Ref. \cite{Ruan2021} was done using monolayer 1T-$\mathrm{TaSe_2}$ on 1H-$\mathrm{TaSe_2}$. They have the same structure as 1T-$\mathrm{TaS_2}$ and 1H-$\mathrm{TaS_2}$. The monolayer 1T-$\mathrm{TaSe_2}$ is also a Mott insulator, while its bulk form is a metal.} Several unusual superconducting behaviors in 4Hb-$\mathrm{TaS_2}$, which may have the origin from this interplay, are observed in experiments. The $T_c$ in 4Hb-$\mathrm{TaS_2}$ is increased to 2.7 K. The two-dimensional nature of the superconducting state is confirmed by the extracted coherence lengths from the upper critical fields. Time-reversal symmetry (TRS) is found to be broken in the superconducting state from muon spin relaxation measurement and is interpreted as a signature for chiral superconductivity. Both $d+id$ and $p+ip$ pairing symmetries that are constrained by the $D_{3h}$ crystal structure are proposed. \cite{Ribak_Skiff2020} The two components superconducting order parameter is further supported by the Little-Parks oscillation and scanning tunneling microscopy and angle-resolved transport experiment. \cite{Almoalem2022} Furthermore, a crossover from the chiral to the nematic state is also detected. \cite{Silber_Mathimalar2022}  

Recent experiments report an unusual magnetic memory effect in 4Hb-$\mathrm{TaS_2}$ single crystals between $T_c=2.7$ K and $T_M=3.6$ K. \cite{Persky2022} Initially, the crystal is cooled below $T=1.7$ K below $T_c$ in a small field $H=1.3$ Oe to create a bunch of randomly distributed vortices due to the pinning potential. Then the crystal is warmed to $T_f>T_c$ under the same field, after which the crystal is zero-field cooled to a target temperature $T<T_c$. Surprisingly, there are randomly distributed vortices despite zero-field cooling. In experiments, this protocol is repeated, but at different $T_f$. The density of the remnant vortices decreases with $T_f$ and disappears when $T_f>T_M$. The remnant vortice density increases linearly with the training magnetic field, and there exists a weak hysteresis near the zero field. The authors of Ref. \cite{Persky2022} also performed annealing from $T>T_M$ to $T_f$ with $T_c<T_f<T_Q$ in the training field. Then the crystal is cooled to $T<T_c$ without a magnetic field. In this process, no field is applied inside the superconducting phase. However, remnant vortices are observed, which is in sharp contrast to the case with zero-field cooling directly from $T>T_M$. The experiments point to an anormalous mangetic memory in $T_c<T<T_M$. In this temperature window, a small magnetization corresponding to one spin in an area of 40 nm$\times$ 40 nm assuming a uniform distribution is observed. This small magnetization is not enough to induce the observed vortex density either due to Zeeman or orbital coupling to the conduction electrons. The authors of Ref. \cite{Persky2022} speculate on the possibility of QSL with breaking TRS residing in the constituting 1T-$\mathrm{TaS_2}$ layers as the origin of the memory effect, without knowing the underlying mechanism. Recently, a $Z_2$ QSL with TRS was proposed to explain the magnetic memory effect, where $Z_2$ vortices or visons are transformed into superconducting vortices at $T<T_c$ upon cooling. \cite{Chen_2022}

\begin{figure}[t]
  \begin{center}
  \includegraphics[width=8.5 cm]{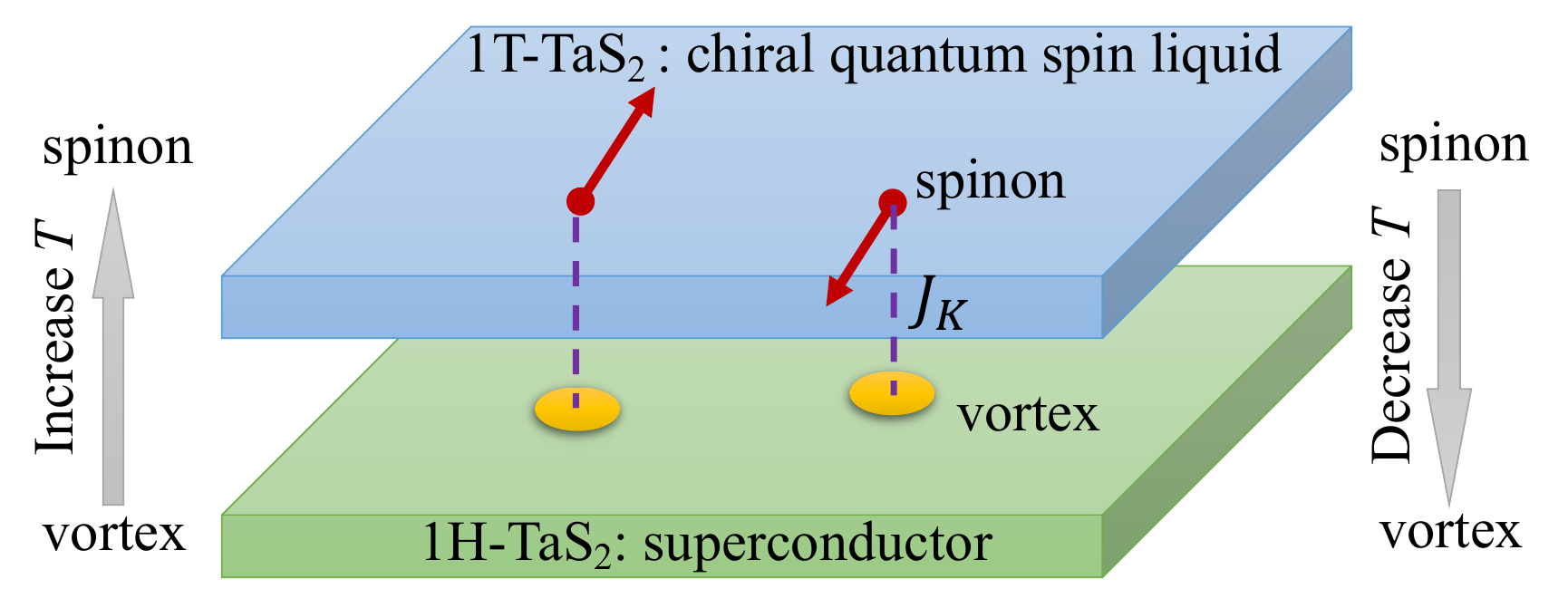}
  \end{center}
\caption{4Hb-$\mathrm{TaS_2}$ is modeled as a multilayer structure with alternating chiral QSL and superconductor layers. Through the interlayer Kondo interaction, a superconducting vortex is dressed by a spinon. Transmutation between spinons and vortices occurs during the thermal cycling process, which results in a magnetic memory effect in the temperature window $T_c<T<T_Q$.  
} 
  \label{f1}
\end{figure}

Our starting point is based on the observation of the Kondo resonance peak in the bilayer of 1T-$\mathrm{TaS_2}$/1H-$\mathrm{TaS_2}$. \cite{Vano2021,Ruan2021} The presence of a Kondo peak suggests negligible charge transfer between 1T-$\mathrm{TaS_2}$ and 1H-$\mathrm{TaS_2}$, and a well-developed localized magnetic moment in 1T-$\mathrm{TaS_2}$ and the existence of itinerant electrons in 1H-$\mathrm{TaS_2}$. Our physical picture is based on the observation that, in the conventional heavy fermion liquid, the Kondo effect converts neutral spinons into charged fermions and, therefore, forms a heavy fermion liquid with an enlarged fermi volume. \cite{PhysRevLett.90.216403} (The charge here is defined under the gauge fields of the electromagnetic fields.) We model 4Hb-$\mathrm{TaS_2}$ as a superconductor-Mott insulator layered structure with an interlayer Kondo coupling, see Fig. \ref{f1}. The Hamiltonian of 4Hb-$\mathrm{TaS_2}$ can be schematically written as $\mathcal{H}=\mathcal{H}_T+\mathcal{H}_K+\mathcal{H}_H(c_i^\dagger, c_i)$, where $\mathcal{H}_T$ ($\mathcal{H}_H(c_i^\dagger, c_i)$) describes 1T-$\mathrm{TaS_2}$ (1H-$\mathrm{TaS_2}$), and $\mathcal{H}_K$ is the Kondo interaction. They have the form (we set $\hbar=e=c=1$ below)
\begin{align}
    \mathcal{H}_T &=\sum_{<ij>} J_{ij}\mathbf{S}_i\cdot \mathbf{S}_j + \sum_{<ijkl>}J_{ijkl} (\mathbf{S}_i\cdot \mathbf{S}_j) (\mathbf{S}_k \cdot \mathbf{S}_l) +\cdots, \\
    \mathcal{H}_K &= J_K \sum_i c_{i,\alpha}^\dagger\mathbf{\sigma}_{\alpha\beta} c_{i\beta}\cdot \mathbf{S}_i.
\end{align}
We have included four spin interactions in $\mathcal{H}_T$ and $\cdots$ represent higher-order interactions. The spin anisotropy in 1T-$\mathrm{TaS_2}$ is small \cite{PhysRevLett.121.046401} and is neglected here. Motivated by the experimental observation \cite{Persky2022} of TRS breaking between $T_{c}$ and $T_m$, and also the identification of chiral QSL in triangular lattice  Hubbard model in the density matrix renormalization group study \cite{PhysRevX.10.021042,PhysRevLett.127.087201}, here we assume that 1T-$\mathrm{TaS_2}$ is in a chiral QSL below $T_Q$. The spin chirality $\mathbf{S}_i\cdot (\mathbf{S}_j\times \mathbf{S}_k)$ with $ijk$ labeling the sites in a  smallest triangle has a nonzero expectation value consistent with TRS breaking.

\begin{figure}[t]
  \begin{center}
  \includegraphics[width=\columnwidth]{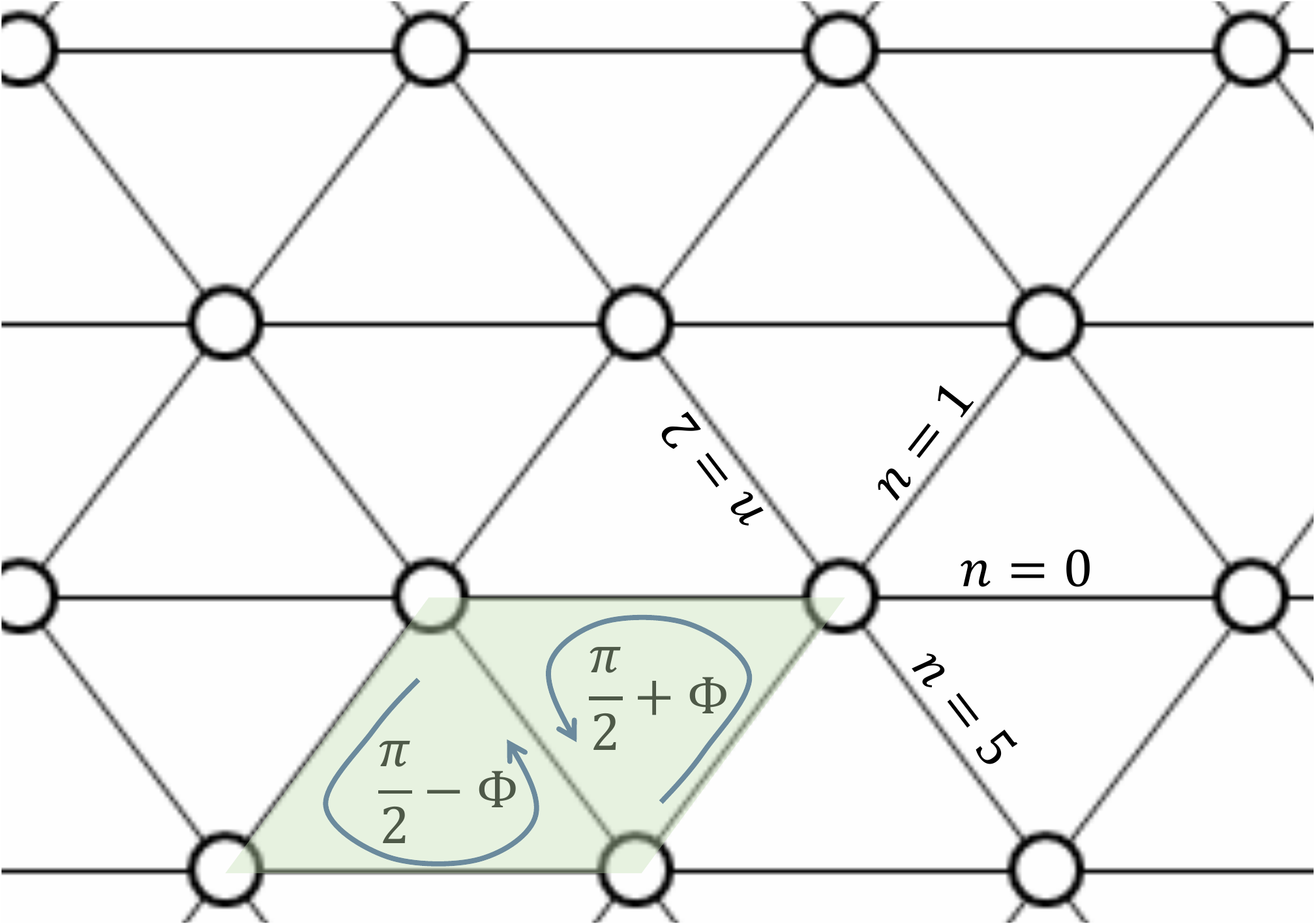}
  \end{center}
\caption{Sketch of flux of the emergent gauge field $\mathbf{a}$ in a triangular lattice. The spinons $f_{\alpha}$ hop in the background of $\mathbf{a}$ flux and form Chern bands. The up and down triangles have $\pi/2-\Phi$ and $\pi/2+\Phi$ flux, respectively. The unit cell is doubled as indicated by the shaded region.} 
  \label{f2}
\end{figure}

The chiral QSL and Kondo coupling can be treated using the parton construction, $\psi_i=(f_{i\uparrow}, f_{i\downarrow}^\dagger)$ with $\mathbf{S}_i=\sum_{\alpha\beta}f_{i\alpha}^\dagger\mathbf{\sigma}_{\alpha\beta} f_{i\beta}/2$ and $f_{i\uparrow}^\dagger f_{i\uparrow}+f_{i\downarrow}^\dagger f_{i\downarrow}=1$. In the mean-field description, $\mathcal{H}_T\approx \sum_{ij} \psi_j^\dagger u_{i,j}\psi_i$. In this construction, $\mathcal{H}_T$ has an SU(2) gauge redundancy \cite{PhysRevB.38.745} and $u_{i,j}$ related by SU(2) gauge rotation are equivalent, i.e. $\psi_i\rightarrow W_i \psi_i$ and $u_{ij}\rightarrow W_i u_{ij} W_J^\dagger$ where $W_i$ is a local SU(2) transformation. \cite{PhysRevB.65.165113}  We will discuss two ansatzes. In the first case, the occupied $f_{i\alpha}$ form a Chern band with Chern number $C=1$, which can be obtained by introducing flux for $f_{i\alpha}$ hopping in the triangular lattice, see Fig. \ref{f2}. The other ansatz corresponds to the state where $f_{i\alpha}$ is in the $d+i d$ superconducting state \cite{PhysRevB.103.165138}.

The Kondo coherence can also be described in terms of the parton theory. \cite{PhysRevLett.90.216403} The Kondo term can be written as $ c_{i\alpha}^\dagger\mathbf{\sigma}_{\alpha\beta} c_{i\beta}\cdot \mathbf{S}_i=-(f_{i\alpha}^\dagger c_{i\alpha})( c_{i\beta}^\dagger f_{i\beta})+\cdots$, where we have neglected a shift in the local chemical potential for $c_{i\alpha}$ fermion. The Kondo coherence emerges when the composite bosons $Q=f_{i\alpha}^\dagger c_{i\alpha}$ condense, $\langle Q\rangle \neq 0$. As a consequence, the SU(2) gauge redundancy in $\psi_i$ is broken down to U(1). $f_\alpha$ fermion carries a charge associated with an emergent gauge field $\mathbf{a}$, which is a subgroup of $W_i$. The $c_\alpha$ fermion carries a physical charge associated with the physical gauge field $\mathbf{A}$ (actual electromagnetic fields). The condensation of $Q$ locks $\mathbf{A}$ to $\mathbf{a}$ as a result of the Higgs mechanism. In the conventional heavy fermion liquid, the $f_\alpha$ fermions become part of the fermi liquid of $c_\alpha$, which enlarges the fermi volume. We ascribe the observed memory effect in 4Hb-$\mathrm{TaS_2}$ to a consequence of the coupling between $\mathbf{A}$ and $\mathbf{a}$ as detailed below. 

The total Lagrangian of the system can be written as
\begin{align}
    \mathcal{L}=\mathcal{L}_T(f_\sigma, \mathbf{a})-\frac{\rho_Q}{2}[\nabla \theta-(\mathbf{a}-\mathbf{A})]^2-\frac{\rho_\Delta}{2}(\nabla \phi-\mathbf{A})^2+\cdots
\end{align}
where we have neglected other terms for simplicity (i.e. quadratic and quartic terms in $Q$ and $\Delta\equiv \langle c_{i\alpha} c_{j\beta}\rangle$). We also neglected the fluctuations in the amplitude of $Q$ and $\Delta$, which are gapped in the ordered state. The kinetic terms describe the coupling between the phase fluctuation of $Q=|Q|\exp(i\theta)$ and $\Delta=|\Delta|\exp(i\phi)$ to the gauge fields. The superconducting order parameter $\Delta$ describes superconductivity in 1H-$\mathrm{TaS_2}$, which can be intrinsic or induced by proximity to the chiral QSL. It is shown that doping chiral QSL can stabilize the chiral $d+id$ superconductivity. \cite{PhysRevLett.125.157002,PhysRevX.12.031009} One may also argue that similar superconductivity can emerge through the Kondo coupling. In our picture, only the Higgs mechanics, which is universal regardless of the pairing mechanism and symmetry, is important for the current discussion. $\mathcal{L}_T$ describes the quantum state of 1T-$\mathrm{TaS_2}$. When $f_\alpha$ form a Chern insulator with the occupied up and down spin bands having a Chern number $C=1$, we can integrate out the $f_\alpha$, which results in the Chern-Simon Lagrangian
\begin{align}\label{eq3}
    \mathcal{L}_T(f_\sigma, \mathbf{a})=\frac{2 C}{4\pi} \epsilon^{\mu\nu\rho} a_\mu\partial_\nu a_\rho-\mathbf{J}_f\cdot \mathbf{a},
\end{align}
where we have included a probe charge $\mathbf{J}_f$. The ground state described by $\mathcal{L}_T$ is doubly degenerate, and supports excitation spinons as semions. Each spinon $f_\alpha$ carries $\pi$ flux of $\mathbf{a}$, which can be seen by varying $\mathcal{L}_T$ with respect to $a_0$. This flux attachment is also responsible for the semion statistics of $f_\alpha$.

One may worry that the flux of $\mathbf{a}$ when locks to $\mathbf{A}$ destroys the superconductivity in the $c_\alpha$ fermions. Eq. \eqref{eq3} corresponds to filling of two spinons per unit flux of $\mathbf{a}$. This can be achieved by doubling the atomic unit cell with each unit cell containing $\pi$ flux, see Fig. \ref{f2}, which amounts to a gigantic magnetic field strength of $\mathbf{a}$. This translates into $\pi$ flux of $\mathbf{A}$ per unit cell when the $Q$ bosons condense. A cooper pair $\langle c_{i\alpha} c_{j\beta}\rangle $ carries $2\pi$ flux of $\mathbf{A}$ per unit cell, which is equivalent to zero flux because the flux is defined modulo $2\pi$ on lattice. Therefore, superconductivity is stable in the presence of $\pi$ flux of $\mathbf{a}$ per unit cell.

The physics depends on three temperatures, $T_c$, $T_Q$ and the Kondo coherence temperature $T_K$. In the experimentally relevant case $T_Q>T_c$, the magnetic memory exists (disappears) when $T_K>T_c$ ($T_K<T_c$). Experimentally $T_K\approx 18$ K in the bilayer of 1T-$\mathrm{TaS_2}$ and 1H-$\mathrm{TaS_2}$ \cite{Vano2021} and we will focus on the case with $T_K>T_c,\ T_Q$ in the following discussion. The transmutation between spinons and vortices depends on the phase stiffness $\rho_Q$ and $\rho_\Delta$. We assume that $\rho_Q\gg \rho_\Delta$ such that $\mathbf{a}$ is always locked to $\mathbf{A}$. In the opposite limit, spinons induce vortices in $Q$ instead.   

Now we are in a position to interpret the magnetic memory effect during the thermal cycling of 4Hb-$\mathrm{TaS_2}$ in experiment. When the system is cooled below $T_c$ under a weak magnetic field, certain magnetic flux is trapped inside the system in the form of Abrikosov vortices and are pinned by defects. Each Abrikosov vortex carries $\pi$ flux of $\mathbf{A}$. The $Q$ condensate locks $\mathbf{A}$ to $\mathbf{a}$ and induces $\pi$ flux in $\mathbf{a}$, which attaches one spinon to the Abrikosov vortex. Warming the system under the same field above $T_c$ but below $T_Q$ destroys Abrikosov vortices. Some of the spinons relax and become part of the Chern bands, while the other can be pinned by defects. When the system is zero-field cooled below $T_c$ from this state, these trapped spinons induce Abrikosov vortices due to the Kondo effect. As the temperature for field training $T_f$ increases toward $T_Q$, the pinning energy of spinons decreases. As a result, the remnant field due to the Abrikosov vortices at the end of thermal cycling decreases with $T_f$. It is also clear that the remnant field increases linearly with the density of the initial Abrikosov vortices, hence the training field as observed in experiment. Hysteresis naturally occurs due to the pinning of spinons and Abrikosov vortices.

The charge of spinon under $\mathbf{a}$ carried by $\pi$ flux of $\mathbf{a}$ depends on the sign of $C$ and hence the sign of $\mathbf{S}_i\cdot (\mathbf{S}_j\times \mathbf{S}_k)$. $\pi$ flux of $\mathbf{a}$ corresponds to positive (negative) charged spinon under $\mathbf{a}$ for $C=1$ ($C=-1$). Positive (negative) charged spinon corresponds to an excess (vacant) spinon in a magnetic unit cell of $\mathbf{a}$. In the mean-field description, vacant spinon can be a localized spin, which does not fractionalize into spinon. An excess spinon corresponds to the case where the magnetic unit cell of $\mathbf{a}$ is enlarged to contain three spin sites. In real systems, it is likely that there exist multiple domains of chiral QSL with different $C$ similar to the Chern insulators. \cite{Wang_Ou2018,Grover2022} In each domain, the charge of the induced spinons depends on $C$. Nevertheless, during the thermal cycling process, the polarization of the Abrikosov vortices is the same as that of the original Abrikosov vortices regardless of the sign of $C$.  

We then discuss the nucleation of Abrikosov vortices for the protocol when the training magnetic field is only applied between $T_c$ and $T_Q$. Without the training field, it is natural to form chiral QSL mosaics with opposite $C$ numbers. Within each domain, there exist regions with excess or vacant spinons. These chiral QSL mosaics with excess or vacant spinons tend to induce Abrikosov vortices and antivortices through the Kondo coupling. Well-separated Abrikosov vortices can form only when the separation of spinons and size of chiral QSL domain far exceed the size of Abrikosov vortices, which is of the order of the London penetration depth. The size of an Abrikosov vortex in 4Hb-$\mathrm{TaS_2}$ is around $5\ \mathrm{\mu m}$  according to the experimental image of vortex using scanning superconducting quantum interference device \cite{Persky2022}. It is likely that the vortex size is comparable to or even larger than the size of the chiral QSL domains. As an example, the size of Chern domains in the Chern insulator is smaller than $1 \mathrm{\mu m}$. \cite{Grover2022} In this case, no Abrikosov vortices are generated during the cooling process without a training field. In the presence of a training field, one type of chiral domain is favored through the weak orbital coupling between the spin chirality $\mathbf{S}_i\cdot (\mathbf{S}_j\times \mathbf{S}_k)$ and the magnetic field. \cite{PhysRevB.73.155115,PhysRevB.78.024402,Banerjee_Lin_2022} As a result, the favored chiral domains grow in size, and the spinons can generate Abrikosov vortices when the domain sizes are larger than the vortex size. The training field also couples to the spinons through a direct Zeeman coupling, but this coupling does not select the polarization of the vortex.

The physical picture presented here suggests the existence of spinons. The spinons in the normal state below $T_Q$ generate a weak magnetization that is much smaller than the magnetization associated with vortices. A weak magnetization corresponding to one spin-1/2 moment in an area of 40 nm $\times$ 40 nm on average is indeed observed in the experiment, \cite{Persky2022} which supports the current physical picture.

We proceed to discuss the other chiral QSL ansatz with $f_\alpha$ in the $d+i \eta d$ pairing state and its connection to the unconventional superconducting properties of 4Hb-$\mathrm{TaS_2}$ observed in experiment. The ansatz for this state is
\begin{align}
u_{\mathbf{r}, \mathbf{r}+\mathbf{r}_n}=\lambda \tau_z + \gamma [ \cos(n 2\pi/3)\tau_x-\eta\tau_y \sin(n 2\pi/3)],    
\end{align}
where $\tau_\mu$ are Pauli matrices acting in the spinor space of $\psi_i$, and $n=0, 1, \cdots, 5$ counterclockwise enumerates the bonds that connect to the site $\mathbf{r}$, see Fig. \ref{f2}. We introduce the parameter $\eta$ to characterize the ratio between the two $d$-wave components. For $\eta\neq 1$, the QSL ansatz breaks the $C_6$ rotation symmetry even in the absence of Kondo coherence when there exists SU(2) gauge redundancy, hence describing a nematic state. This can be seen by writting  $u_{\mathbf{r}, \mathbf{r}+\mathbf{r}_n}=\mathbf{d}_n\cdot \mathbf{\tau}$. $|d_n|$ is different for different bonds. SU(2) gauge transformation preserves $|d_n|$ therefore cannot restore the $C_6$ rotation symmetry.  The chiral QSL with the $d+i \eta d$ ansatz has $Z_2$ gauge symmetry, and supports $e$, $m$ and $\epsilon$ anyons. \cite{PhysRevB.95.014508} The $m$ anyon is a vortex of the $d+i \eta d$ pairing state and carries $\pi$ flux of $\mathbf{a}$. The discussion of the transmutation between spinons and Abrikosov vortices for the Chern band ansatz can be extended straightforwardly to the chiral QSL with $d+i \eta d$ ansatz, but with the replacement of spinons by $m$ anyons. The $d+i \eta d$ ansatz is relevant here because it automatically engenders chiral superconductivity with $d+i \eta d$ pairing symmetry when Kondo coherence sets in. This provides a mechanism for the observed chiral and nematic superconductivity in 4Hb-$\mathrm{TaS_2}$ with elevated $T_c$ compared to the 2H-$\mathrm{TaS_2}$ compound. \cite{Ribak_Skiff2020,Silber_Mathimalar2022}

{\it Discussion.---} The related idea of the transmutation between visons in $Z_2$ QSL and Abrikosov vortices was discussed in the context of cuprate superconductors. \cite{PhysRevLett.86.292} There the pseudogap phase is identified as a $Z_2$ QSL, which can be accessed by raising the temperature from the $d$-wave superconducting state. In cylindrical geometry, a trapped magnetic flux in the superconducting state is transmutated to a $Z_2$ vison or vice versa during the thermal cycling process. This was proposed as smoking-gun evidence for the existence of topological order in the pseudogap phase. This idea is adopted and extended by Chen \cite{Chen_2022} to explain the magnetic memory effect in 4Hb-$\mathrm{TaS_2}$. The proposed $Z_2$ QSL preserves TRS, which contradicts the experimentally observed TRS breaking state above $T_c$. In experiments, when the system is cooled from the field trained state above $T_c$, only Abrikosov vortices with polarization aligned with the training field are generated. Here, we assume a chiral QSL, and polarization of the magnetic field is memorized by the system through the Chern number of the spinon bands or sign of $\mathbf{S}_i\cdot (\mathbf{S}_j\times \mathbf{S}_k)$. The polarization of the generated Abrikosov vortices therefore is determined by the training field, which is consistent with the experiments. Furthermore, the chiral QSL can also provide a mechanism to stabilize chiral superconductivity in 4Hb-$\mathrm{TaS_2}$.

We remark that the U(1) QSL, where spinons form a neutral Fermi surface and are coupled to an emergent U(1) gauge field, was proposed for 1T-$\mathrm{TaS_2}$. The QSL can persist upto 200 K. It is possible that the coupling between 1T-$\mathrm{TaS_2}$ and 1H-$\mathrm{TaS_2}$ in 4Hb-$\mathrm{TaS_2}$ modifies the magnetic interactions and stabilizes the chiral QSL below $T_Q$. There can be a transition between the U (1) QSL and the chiral QSL. A finite temperature transition of chiral QSL is allowed even in the two-dimensional limit because the QSL breaks the discrete $Z_2$ symmetry associated with $\mathbf{S}_i\cdot (\mathbf{S}_j\times \mathbf{S}_k)$. The chiral QSL in Mott insulator generates quantized thermal Hall conductance. In the metallic 4Hb-$\mathrm{TaS_2}$, there can exist a chiral metallic state in the temperature region $T_c<T<T_Q,\ T_K$, which can be verified by thermal or electrical Hall conductance measurement. This chiral metallic state may also be absent when superconductivity is induced by the Kondo coupling to the chiral QSL with the $d+i\eta d$ ansatz, such that $T_c=T_K$.

It is known that the Kondo effect connects the QSLs to their charged partners through the Higgs mechanism associated with the Kondo coherence. \cite{PhysRevLett.90.216403,Hsieh_Lu_Ludwig_2017} As a concrete example, here we demonstrate the transmutation of neutral excitations to charged excitations under physical electromagnetic fields in the chiral QSL and superconductor heterostructures. This points to a promising route for experimental detection of neutral excitations in strongly correlated systems, which is at the heart of the controversy in modern condensed matter physics. Therefore, the 1T-$\mathrm{TaS_2}$ and 1H-$\mathrm{TaS_2}$ heterostructures are a promising platform to demonstrate this physics by varying $T_c, T_K$ and $T_Q$ in a controlled way.

In summary, we show that the Kondo coupling between a chiral quantum spin liquid and a superconductor allows transmutation between spinons and superconducting vortices. This results in a magnetic memory effect, which explains the same effect observed in 4Hb-$\mathrm{TaS_2}$. Our results further support the existence of the chiral quantum spin liquid in 4Hb-$\mathrm{TaS_2}$. We also suggest that the chiral quantum spin liquid may be responsible for the unconventional superconductivity in 4Hb-$\mathrm{TaS_2}$.

\begin{acknowledgements}
{\it Acknowledgments.---} The work at LANL was carried out under the auspices of the U.S. DOE NNSA under contract No. 89233218CNA000001 through the LDRD Program, and was performed, in part, at the Center for Integrated Nanotechnologies, an Office of Science User Facility operated for the U.S. DOE Office of Science, under user proposals $\#2018BU0010$ and $\#2018BU0083$.

\end{acknowledgements}

\bibliography{references}

\end{document}